\documentclass[aps,reprint,superscriptaddress,noeprint,amsmath,amssymb,prb,showkeys,floatfix,nofootinbib]{revtex4-2}

\usepackage{epsfig}
\usepackage{times}
\usepackage{color}
\usepackage{graphicx}
\usepackage{float}
\usepackage{amsmath}
\usepackage[caption=false]{subfig}
\usepackage{appendix}
\usepackage[hidelinks=true,colorlinks=true,linkcolor=blue,citecolor=blue]{hyperref}

\begin{document}

\title{A rule-free workflow for the automated generation of databases from scientific literature}
\author{Luke P.~J. Gilligan$^\dag$}

\affiliation{School of Physics, AMBER and CRANN Institute, Trinity College, Dublin 2, Ireland}
\author{Matteo Cobelli$^\dag$}
\affiliation{School of Physics, AMBER and CRANN Institute, Trinity College, Dublin 2, Ireland}

\author{Valentin Taufour}
\affiliation{Department of Physics and Astronomy, University of California, Davis, California 95616, USA}

\author{Stefano Sanvito}
\email[Corresponding Author: ]{sanvitos@tcd.ie}
\affiliation{School of Physics, AMBER and CRANN Institute, Trinity College, Dublin 2, Ireland}

\begin{abstract}
In recent times, transformer networks have achieved state-of-the-art performance in a wide range of natural language processing tasks. Here we present a workflow based on the fine-tuning of BERT models for different downstream tasks, which results in the automated extraction of structured information from unstructured natural language in scientific literature. Contrary to existing methods for the automated extraction of structured compound-property relations from similar sources, our workflow does not rely on the definition of intricate grammar rules. Hence, it can be adapted to a new task without requiring extensive implementation efforts and knowledge. We test our data-extraction workflow by automatically generating a database for Curie temperatures and one for band gaps. These are then compared with manually curated datasets and with those obtained with a state-of-the-art rule-based method. Furthermore, in order to showcase the practical utility of the automatically extracted data in a material-design workflow, we employ them to construct machine-learning models to predict Curie temperatures and band gaps. In general, we find that, although more noisy, automatically extracted datasets can grow fast in volume and that such volume partially compensates for the inaccuracy in downstream tasks.
\end{abstract}

\maketitle

\def\thefootnote{$\dag$}\footnotetext{Equal contribution.}

\section{Introduction}
Since the dawn of modern science, there has been a continuous, exponential growth in the volume of published scientific literature~\cite{Bornmann2021}. When related to materials science, such an abundance of data clearly offers a wide range of possibilities and opportunities. Materials data, in fact, can provide the foundation for models and theories to navigate the physical/chemical space and ultimately drive discovery. Unfortunately, access to information from unstructured literature at such a massive scale presents significant technical and practical challenges. As a result, in general, curated databases are scarce and often limited to theoretical data only. This is because for theory data one does not need to exploit the literature, but simply to run highly standardised first-principles calculations, which are amenable to automated collection~\cite{CURTAROLO2012227,Talirz2020,materials_project,OQMD}. Importantly, large-scale theoretical datasets have been proven to be a revolutionary tool in the search for new materials with unique properties and for the discovery of intricate materials trends. For instance, they have been used to predict the existence of novel magnets~\cite{stefano}, to identify materials regions favourable to superconductivity~\cite{Isayev2015}, to design novel high-entropy alloys~\cite{LEDERER2018364}, to identify low-thermal-conductivity compounds~\cite{PhysRevX.4.011019}, or to predict the $zT$ thermoelectric figure of merit in inorganic materials~\cite{Xi2018}, just to name a few examples. Furthermore, theory datasets have been a platform for constructing machine-learning (ML) models with enhanced throughput~\cite{PhysRevLett.114.105503,Isayev2017,doi:10.1021/acs.chemmater.9b05342}.

Although no calculation can replace an experiment, databases containing experimental results are much rarer and typically smaller. In general, these are extremely labour-intensive to generate, since the information is not directly collated at the laboratories but needs to be mined from the published literature. As a consequence, experimental databases are not comprehensive but usually list only a handful of properties for each compound, such as the crystal~\cite{COD,ICSD,groom2016cambridge} or magnetic structure~\cite{Bilbao}. Furthermore, they update slowly and, given the effort needed to construct them, they are often proprietary in nature, despite the efforts of some open-access initiatives~\cite{COD,Bilbao}. Thus, the existing landscape of experimental datasets is incomplete and fragmented, and most of the known experimental results remain accessible only through unstructured scientific literature. 

This is a serious drawback for materials innovation, not only because experimental data corresponds to reality, but also because experimental data may often contain information inaccessible to simulations or may be located where simulations are prohibitive or highly inaccurate. It is also important to note that typical ML models for property predictions are often very data-hungry so the absence of large datasets of experimental data precludes their formulation. Still, there have been a few successful examples where only experimental data was used to construct ML predictors and classifiers for materials properties~\cite{James_tc, Tc_super_cond,Zhuo2018}, and examples of transfer learning between theoretical and experimental data~\cite{Chen2021}. These suggest that the creation of structured experimental databases may represent a significant asset in materials design.

Given the large volume of literature regularly published, this appears as a daunting task. In order to understand the scale of the problem, consider that there are 231 journals listed in the materials science and engineering category of the Clarivate Master Journal List (mjl.clarivate.com/home). If the average number of articles published yearly in a journal is around 1,000, we will have about a quarter of a million articles with materials' information published per year. Clearly, experimental databases of such capacity cannot be curated with laborious manual methods, indicating that automatisation is the only way possible. The current state-of-the-art method for automatizing the data-extraction process from literature is ChemDataExtractor~\cite{ChemDataExtractor}. ChemDataExtractor relies upon rules-based text parsing, coupled with conditional random fields for named entity recognition (NER)~\cite{crf}. Hence, the performance of a new model depends on the ability of the user to adequately define rules. Furthermore, since different quantities can be described by natural language in structurally different ways, every new extraction task requires the user to define new grammatical and syntactic rules. As a result of these features, the process is still labour-intensive and reduces our ability to conduct the large-scale deployment of such methods.

In recent years, significant progress has been made in constructing tools that improve our ability to automate further this extraction procedure. Textual representations that are suitable for advanced, context-aware, natural language processing (NLP) have long been a focus of research in this domain. Simple representations such as one-hot encoding of dictionaries of vocabulary, bag of words models, or statistical representations weighting the importance of certain terms over others, such as term frequency-inverse document frequency (TF-IDF)~\cite{SPARCKJONES1972}, have been the go-to representations for many of the more conventional NLP tasks. For instance, they have been successfully used in sentiment analysis or simple classification tasks. These representations are still adequate for handling large-scale texts with simple models. However, in order to operate on a level at which the extracted individual terms or sentences are processed accurately, we must turn to more sophisticated methods of representation.

Word embedding naturally follows as a viable candidate for such applications. Word embeddings are textual representations that aim to describe words in a vocabulary as vectors in a high-dimensional vector space. In this space, the similarity between words is captured by the projection of one such vector onto the other. There are numerous algorithms to learn embeddings from a corpus of documents. Examples of the most widely-used algorithms are word2vec~\cite{word2vec} and GloVe~\cite{pennington-etal-2014-glove}, models that have been also used for scientific text embedding. For instance, the word2vec algorithm was used by the Materials Genome Initiative to train a domain-specific materials-science embedded representation~\cite{Tshitoyan2019}. This embedding was demonstrated to exhibit a good knowledge of the chemical space and insights into the physical properties of materials could even be extracted from purely text-based representations trained on scientific literature alone.

Transformer networks elaborate on these ideas and are the current state of the art in the representation of natural language~\cite{vaswani}. The core concept of a transformer network is the use of self-attention to capture the syntactic interdependencies between words, meaning that these networks exhibit a superior ability to parse the context in which a term appears in a sentence. Arguably, one of the most prevalent transformer-based models for NLP is the Bidirectional Encoder Representations from Transformers (BERT), a large-scale architecture with approximately hundreds of millions of tuneable parameters.\cite{devlin-bert} Since its conception in 2018, BERT has rapidly become the language platform for models in many NLP applications, achieving state-of-the-art performance across a range of benchmarks~\cite{bert_benchmark,albert}. Furthermore, there have been a number of BERT architectures adapted to outperform the conventional BERT model in various disparate NLP domains, including but not limited to the fields of general scientific literature~\cite{scibert}, financial sentiment analysis~\cite{finbert}, and biomedical text-mining~\cite{Lee_2019}. There is also a previously fine-tuned architecture for the field of materials science, called MatSciBERT~\cite{Gupta2022}, which is one of the pre-trained architectures we have employed in this work.

Currently, much focus is being placed on the use of autoregressive large language models, as a means of achieving state-of-the-art performance in a large variety of natural language tasks. While these models are giving promising results there are some significant drawbacks associated with their use. Firstly, their size imposes heavy hardware requirements at inference time and even larger requirements for fine-tuning. It is possible to use services provided by private companies, such as OpenAI, to have access to these models through API. However, the fees can become prohibitive for the large number of prompts needed by an information-extraction workflow. It is also worth considering that for such generative large language models, there is little to no ecosystem of domain-specific pre-training, a feature of the BERT architectures. Indeed, non-domain-specific generative models have not been shown to be superior to domain-specific BERTs in the field of material science. Finally, generative models are known to hallucinate, meaning that they currently have a tendency to make up false information in the generated text. When processing large amounts of data, it is impossible to verify that a generative model has not manufactured some of the data in the resulting database. This is a particularly critical problem in the context of the automatic generation of databases since each entry is required to be reliably linked to its source. Common strategies to limit the effects of hallucinations include iteratively prompting to double-check the output. However, these approaches rapidly increase the number of tokens passed as prompt, hence the computation time or the API access cost for each extraction. For these reasons, we believe that an information-extraction workflow based on domain-specific BERT models can be useful for the community since it can be deployed and fine-tuned on commonly available hardware. 

In this work, we introduce a rule-free workflow for the automatic extraction of information from scientific literature. The extraction is performed by means of a sequence of BERT models finely tuned on specific downstream tasks. The superior contextual awareness of the BERT representation allows the necessary grammatical and syntactic rules for extraction to be learnt by the transformer model from a sample of labelled text. The text-labelling step now substitutes the design of a rule-based grammar and it does not require any previous knowledge of natural language processing or coding to develop new extraction procedures. This is enabled by our self-contained literature-to-structured-properties database pipeline, which is here named BERT Precise Scientific Information Extractor (BERT-PSIE). Moreover, by leveraging the transfer-learning capabilities of BERT models, the convergence in performance on the downstream tasks is reached with a relatively small training set of labelled text. Each entry of the database generated by BERT-PSIE can be linked to a specific source without any possibility of hallucination since all the language models used are trained for classification tasks and not used in a generative setting.

Value has been already demonstrated for databases automatically generated with previously developed schemes, providing useful insights into a range of physical and chemical properties of compounds~\cite{Kim2017,EKim2017,Nandy2021}. More recently, there have been also several inroads made in the development of transformers for materials science applications~\cite{Huang2022,matextract}. However, all these attempts present a common shortcoming. Namely, when validating the quality of the dataset automatically generated by NLP methods, one does not have available a reference set of manually curated entries to compare with. This means that the automatically generated datasets cannot be tested against an established ground-truth reference. This issue is addressed here by choosing the properties for which we can obtain a manually curated dataset. The first of these properties is the Curie temperature, $T_\mathrm{C}$, for which we avail of the combined manually-curated databases from Nelson \textit{et al.}~\cite{James_tc} and Byland \textit{et al.}~\cite{Valentin-co}, and of one automatically generated using ChemDataExtractor~\cite{snowball}. The second property of interest is the electronic band-gap, for which we have a manually-curated dataset aggregated by Zhuo \textit{et al.}~\cite{BandGapData}. We will show that with a modest amount of labelled data, it is possible to train models that have performance on par with the state-of-the-art rule-based methods. Furthermore, we demonstrate that the automatically generated data can be used to construct both $T_\mathrm{C}$ and band-gap predictors, such that they can already be used for predicting materials properties.

\section{Results \& Discussion}

\subsection{BERT-PSIE}

Results for the entire workflow are now presented. In each of the following sections, we first examine the BERT-PSIE pipeline fine-tuned for the Curie temperatures extraction task, followed by its counterpart, which is specialised for electronic band gap extraction. We begin by discussing the performance (precision, recall and $F_1$ score) achieved on a test set by each of the three main modules composing our extraction workflow. This is followed by a direct comparison with ChemDataExtractor, where we evaluate the information extracted by both workflows from the same set of abstracts. Finally, we employ the BERT-PSIE pipelines to perform information extraction from a corpus of scientific papers and compare the resulting automatically generated database with our manually curated ones~\cite{James_tc,Valentin-co,BandGapData} and those created with ChemDataExtractor~\cite{snowball, Dong2022}. Additionally, we evaluate the performance of the $T_\mathrm{C}$ database in the screening for magnetic compounds by training ML models to predict the property in question.


\subsection{Fine-Tuning of BERT Models}

\begin{table}[h]
\centering
\caption{Performance of the three modules developed for the Curie temperature extraction: the sentence-level relevancy classifier, the NER and the relation classifier. Results are presented for the test sets. Here we report: precision, $P$, recall, $R$, and $F_1$ score. The size of the test (TeS) and training (TrS) sets are also given (number of sentences used). For the case of NER we report results for both chemical entities 
(Chem) and $T_\mathrm{C}$, as well as the support.}\label{table:sentence_class}
\begin{tabular}{c | c c c c c | c c}\hline\hline
Model & Entity & $P$ & $R$ & $F_1$ & Support & TrS & TeS \\ 
 \hline
Classifier &  & 0.83 & 0.80 & 0.81 & & 3941 & 801 \\ \hline
NER & Chem & 0.92 & 0.86 & 0.89 & 754 & 1,769 & 168 \\   
 & $T_\mathrm{C}$ & 0.97 & 0.81 & 0.88 & 42 &  &  \\ \hline
Relation &  & 0.72 & 0.64 & 0.68 &  & 200 & 50 \\ \hline\hline
\end{tabular}
\end{table}

In this section we examine the performance of the BERT-PSIE modules in isolation. These modules consist of BERT models fine-tuned for different downstream tasks, namely, sentence classification, NER and relation classification. 

Let us discuss the Curie-temperature extraction task first. The evaluation metrics for the sentence-level relevancy classifier are presented in the upper row of Table~\ref{table:sentence_class}. Precision, $P$, and recall, $R$, are both above 0.8, indicating high-level model performance on the test data. Care was taken to ensure that the training data was as representative as possible of the literature. However, similarities in the syntactic structure reporting a temperature value are unavoidable, a fact that increases the level of noise in the extracted data. For example, consider the sentence `Barium titanate (BaTiO$_3$) is a ferroelectric with a Curie temperature of 120~$^\circ$C'. In this case `Curie temperature' refers to a paraelectric-ferroelectric transition and not to ferromagnetism, but the syntactic structure is almost identical to what was found when describing the magnetic $T_\mathrm{C}$. Such ambiguity can also be found in construction such as `The melting temperature of a compound marks the solid-liquid phase transition. This critical temperature for Fe is 1,538~$^\circ$C'. Since the classification is performed at the sentence level, the content of `This critical temperature for Fe is 1,538~$^\circ$C' is evaluated independently from that of the preceding sentence. Then, it is expected to be erroneously classified. These limitations are inherent to working at the sentence level and further work needs to be done in order to resolve the issue effectively. Here, we mitigate the drawback by selectively analysing scientific texts taken from the magnetism subject area.

The second row of Table~\ref{table:sentence_class} presents the results for the named-entity-recognition step of our automated extraction pipeline. The precision, recall and $F_1$ score of both the classified entities, compound and Curie temperature, are all consistently very high, indicating an excellent performance of the BERT model for token classification. This allows us to extract mentions of compounds and Curie temperatures from the sentences identified using the sentence-level relevancy classifier. Furthermore, given the context-aware nature of BERT-based language models, similar entities can be discriminated against, based on the syntactic and grammatical context in which they appear. In practice, our BERT model can, for the most part, differentiate between temperatures generally and those specifically mentioning critical temperatures. It should be noted, however, that this context-awareness suffers the same weakness as mentioned above in being limited in its ability to differentiate between different types of critical temperatures relating to phase transitions. Regardless, it performs adequately for the purpose of recognising both compounds and Curie temperatures.

Relationship extraction has proved to be the most challenging task in our pipeline due to the sheer quantity of potential combinations of words in various syntactic structures. This also constitutes the major challenge when defining rule-based grammar constructions for methods such as ChemDataExtractor. The last row of Table~\ref{table:sentence_class} summarizes the key evaluation metrics for the BERT relationship-extraction model. Although this is the module presenting the lowest scores, the model still exhibits reasonably good performance and, therefore, it is useful to associate the correct compound-property pairs, thus improving the quality of the final database. In the following sections, different alternative schemes for associating compounds and properties are compared with this relations classifier system. Furthermore, we will discuss the effect of considering Curie-temperature values taken from phrases with multiple compound-property mentions on the integrity of the final extracted database. 

\begin{table}[h]
\centering
\caption{Performance of the three modules developed for the band-gap extraction: the sentence-level relevancy classifier, the NER and the relation classifier. Results are presented for the test sets. Here we report: precision, $P$, recall, $R$, and $F_1$ score. The size of the test (TeS) and training (TrS) sets are also given (number of sentences used). For the case of NER, we report results for both chemical entities 
(Chem) and band gap, as well as the support.}\label{table:band_sentence_class}
\begin{tabular}{c | c c c c c | c c}\hline\hline
Model & Entity & $P$ & $R$ & $F_1$ & Support & TrS & TeS \\ 
 \hline
Classifier &  & 0.95 & 1.00 & 0.97 & & 404 & 134 \\ \hline
NER & Chem & 0.80 & 0.96 & 0.87 & 1166 & 4000 & 1000 \\   
 & Band-Gap & 0.78 & 0.97 & 0.87 & 119 & & \\ \hline
Relation &  & 0.88 & 0.88 & 0.88 & & 300 & 80 \\ \hline\hline
\end{tabular}
\end{table}

The same workflow is then executed for the case of the electronic band-gap extraction task, and the performance metrics of each of the steps in our pipeline are summarized in Table \ref{table:band_sentence_class}. The performance of the sentence-level classifier, first row, is indeed excellent on the test set, with a perfect recall and a slightly lower precision at 0.95. These metrics are even higher than those found for the Curie temperature and indicate an almost perfect ability in differentiating between sentences that contain or do not contain information about the band-gap of a compound. It is assumed that the superior performance of this classifier, even with fewer training examples, is due to the reduction in ambiguity in reporting band gaps, when compared to the likely similarities in syntactic structures that result from the reporting of critical temperatures. Similarly, the NER step consistently remains very performant, with a high precision, recall and $F_1$ score. Finally, we find that the relationship-extraction step for the case of the band-gap outperforms significantly that of the Curie temperature. This is believed, once again, to be related to the more consistent way in which band gaps tend to be reported within the scientific literature, with the sentence structures generally more formulaic than in the case of Curie temperature mentions. This hypothesis is further highlighted by the improved performance of extraction methods relying on the rule associating compounds with band gaps based on the order in which they appear in the sentence. This enhancement in performance is discussed further in section \ref{sec:DataQuality}.

\subsection{Comparison with Rule-Based Methods}

\begin{table}[h]
\centering
\caption{Performance of the extraction carried by BERT-PSIE and the rule-based ChemDataExtractor performed on the same corpus. The comparison is executed on 200 annotated abstracts for each one of the tasks, namely the $T_\mathrm{C}$ and band-gap extraction. The precision, recall and $F_1$ score are presented for BERT-PSIE (single mentions only and the full pipeline), ChemDataExtractor and the combination of the two methods. The manually annotated datasets have a support of 45 entries for $T_\mathrm{C}$ and 109 entries for the band gap.}\label{table:direct_comp}
\begin{tabular}{c | c| c |c }\hline
Model & $P$ & $R$ & $F_1$ \\ 
\hline \hline
\multicolumn{4}{c}{Curie Temperature}  \\ \hline \hline
ChemData. & 0.67 & 0.49 & 0.56  \\ \hline
Single Mentions & 0.82 & 0.20 & 0.32 \\ \hline
BERT-PSIE & 0.67 & 0.31 & 0.42 \\ \hline
BERT-PSIE + ChemData. & 0.64 & 0.64 & 0.64 \\ \hline\hline
\multicolumn{4}{c}{Band-Gap}  \\ \hline \hline
ChemData. & 0.68 & 0.55 & 0.61 \\ \hline
Single Mentions & 0.78 & 0.23 & 0.35  \\ \hline
BERT-PSIE &  0.70 & 0.40 & 0.51  \\ \hline
BERT-PSIE + ChemData. & 0.63 & 0.72 & 0.67 \\ \hline
\end{tabular}
\end{table}

To directly compare the performance of our BERT-PSIE pipeline with the state of the art of rule-based methods we follow a similar approach as designed in \cite{snowball, Dong2022}. A total of 200 abstracts were manually annotated for the case of compound-Curie temperature extraction and separately for compound-band-gap extraction. To obtain this corpus of test samples, unique abstracts were taken, once again, from the arXiv dataset \cite{arXivData}. It should be noted that care was taken to guarantee that there was no overlap between the abstract used for the fine-tuning or validation of the model and the ones used in these test sets. The new test abstracts were obtained by running a keyword search on a sample of unused abstracts in the database. For the Curie temperature test, the keyword search was `Curie', but excluded abstracts containing the term `Weiss'. The band-gap test corpus was constructed using a keyword search for abstracts containing any of the terms `band gap', `band-gap' or `bandgap', alongside the term `eV'. Random abstracts were sampled from the resulting corpus to give a test dataset of 200 for both cases. This selection aimed to increase the number of positive extraction targets (the support). 

Both BERT-PSIE and ChemDataExtractor were run on this test set. A record in the extracted database was deemed true positive only if all entities in the target compound-property pair were present and matched the manual annotation. The number of true positives, false positives and false negatives for the extraction tasks were manually counted for each property extraction, allowing for the calculation of the precision, recall and $F_1$ score for each model. The ChemDataExtractor model for the extraction of Curie temperature came from the same rule-based pipeline as in Ref.~\cite{snowball}. It was not possible to include the snowball model of this extraction pipeline to make it fully hybrid, as the model used in the work was not readily available. In the case of the band-gap extraction with ChemDataExtractor, however, the full hybrid extraction was performed using the model from Ref.~\cite{Dong2022}. The results of this comparison are presented in Table \ref{table:direct_comp}. 

As can be seen from the presented results, the precision of both models is very consistent across property extractions, with the full BERT-PSIE pipeline even slightly outperforming the hybrid ChemDataExtractor model in the case of the band-gap extraction. To isolate the impact of the relationship extraction module and estimate the amount of noise that it introduces, we also consider the case where we only extract from sentences containing single compound-property relationships. We denominate this case as `Single Mentions' and we observe for this case a significant increase in the precision of the extraction. However, this gain in precision is compensated for by a severe reduction in the recall, thereby reducing the overall $F_1$ score. The BERT-PSIE pipeline results in being more selective in the data extracted as any potential contextual ambiguity in the reporting of the value is far more likely to deter a context-based system than a rules-based one. This fact results in a general decrease in the recall when compared to the rules-based and hybrid pipelines. Moreover, one could argue that recall is generally a less important value for the fidelity or usefulness of the resulting database. To expand on this point we define in this work metrics that try to simulate real use cases of the extracted data with the goal of evaluating the usefulness of the resulting automatically generated database. This discussion is reported in sections \ref{sec:DataQuality} and \ref{sec:DataQualityBG}.

The final evaluation, for both extraction targets, was performed on the combination of the extracted values with BERT-PSIE and ChemDataExtractor, resulting in a sharp increase of the recall for the combined dataset, when compared with either of the other methods. The implication of this result is that the rules-based pipeline and the BERT-based pipeline have different strengths in extracting quantities relative to each other, as the distinct increase in recall implies that there is quite a little overlap in the quantities extracted in either case. The reduction in precision relative to either of the two pipelines for this particular test is, due to the fact that the true positive extractions that overlap for the two pipelines are not double counted, but any incorrectly extracted values are added to the total database, leading to an increase in false positives relative to the true positives, and therefore a reduction in the total precision relative to either constituent database.

\subsection{Extraction Performance \& Database Structure}
\begin{figure*}[ht!]
    \centering
    \includegraphics[width=\linewidth]{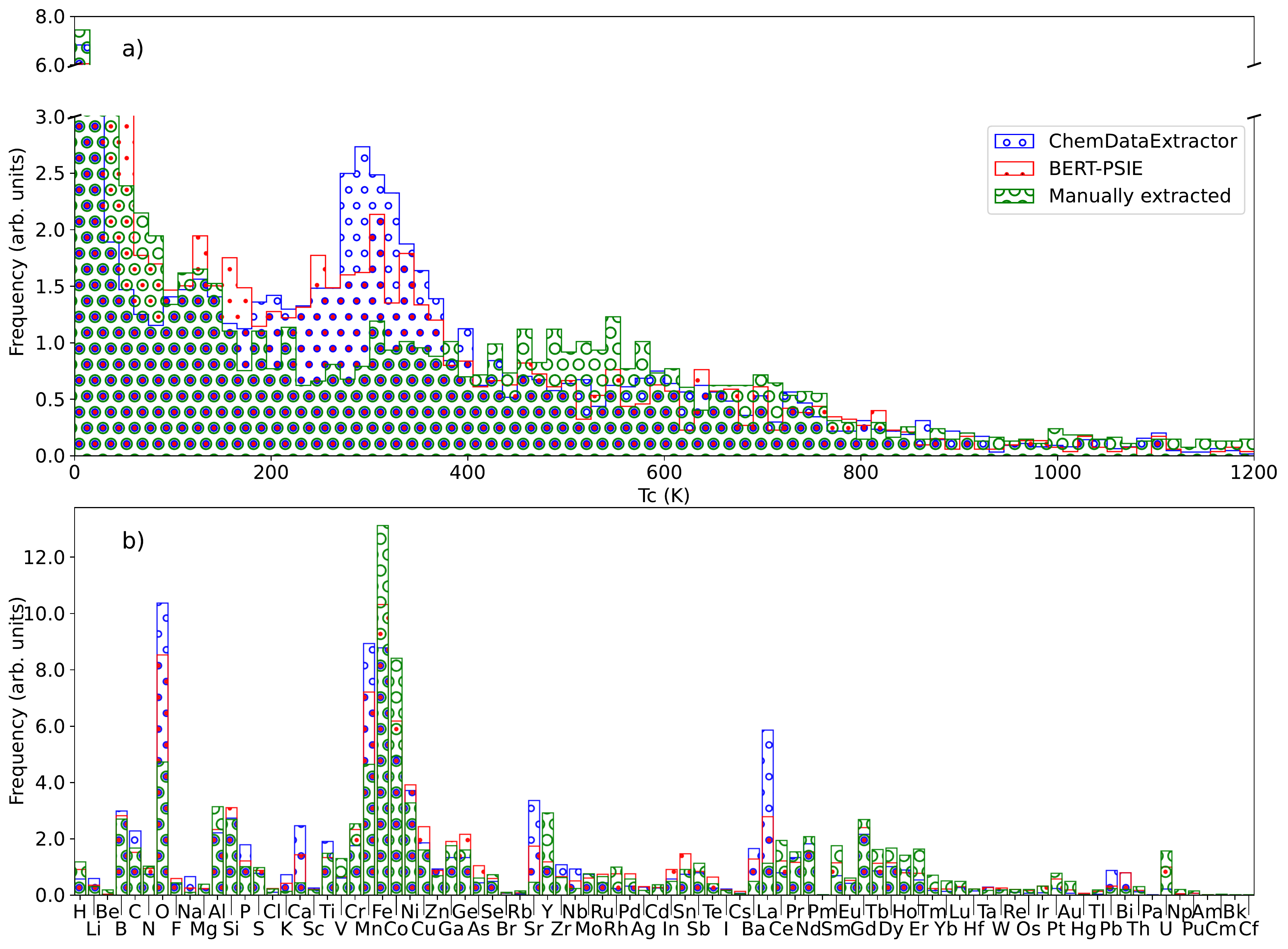}
    \caption[Comparison between the data distributions of the Curie temperature databases.]{Comparison between the content of the different databases: \textit{(red box)} BERT-PSIE, \textit{(blue box)} ChemDataExtractor and \textit{(green box)} the manually-extracted database of Ref.~[\onlinecite{James_tc}]. \textbf{(a)} Normalized distribution of the Curie temperatures extracted. A peak is visible in the distribution around 300~K in both the autonomously extracted databases, which is not seen in the manually extracted one. \textbf{(b)} Relative elemental abundance across the compounds present in a database. Although there is general agreement among the three databases, additional peaks are observed for various elements in the case of automatically extracted data, which are not present in the manually curated dataset. The most severe of these discrepancies is in the relative abundance of Mn- and O-containing compounds. Note that the automatically extracted datasets and the manually curated one are based on different literature libraries.}
    \label{fig:distro}
\end{figure*}

For this section, our discussion begins with the case of the Curie temperature extraction. Approximately 180,000 full-text XML papers were downloaded and split into lists of sentences. These sentences were subsequently run through the sentence-relevancy classifier, yielding a total of approximately 55,000 sentences considered relevant. Here a sentence is considered relevant if it is likely to contain a Curie temperature mention. The resulting sentence list was passed through the named entity recognition system (NER-BERT). Sentences containing a single mention of a Curie temperature and a single compound were directly added to the database. When the sentences contained multiple references to compounds and/or Curie temperatures, then a list of sentences with all the possible pairs of entity mentions was built. These were then classified with the relationship classification BERT model. The compound/property pair predicted by the model to be correct is then added to the database.

The data extracted is subsequently post-processed by scaling all the temperatures to units of Kelvin and by filtering out compositions that could not be expressed as a combination of chemical elements (i.e. commercial or colloquial names for compounds). All chemical formulas are scaled to have normalized integer coefficients (e.g. Ga$_{0.5}$Fe$_{2.5}$O$_4$ becomes GaFe$_{5}$O$_{8}$).

After post-processing, the final extraction gave us a database containing 3,518 distinct compound-property entries together with their digital object identifiers (DOI).

For the case of the band-gap extraction, approximately 77,000 papers were downloaded and split into lists of sentences. Running the sentence-level classifier yielded a dataset of approximately 126,000 sentences deemed likely to contain a band gap. The same strategy as above was adopted for the NER and the relationship classification steps. Finally, after the post-processing step, expressing all units as eV, a final database of 2,090 unique compound-property relationships was yielded.

Evaluating the metrics (precision, recall and $F_1$ score) of a data-extraction method provides only limited information on its performance in that it gives only general indications of the quality. Clearly, the ultimate test is set by its success in the extraction task it has been designed for, which is by the quality of the data extracted and by their potential use in downstream tasks (e.g. the construction of ML models). Performing such an assessment is generally challenging since one lacks manually curated data that are difficult to assemble because of the elevated time investment involved. In our case, the situation is much more favourable, since we can compare our automatically extracted data with manually-curated databases from various sources. 

In the case of the Curie temperatures, we avail of the database of Nelson \textit{et al.}~\cite{James_tc}. This dataset has been created by aggregating the \textit{AtomWork} database~\cite{Yamazaki2011}, \textit{Springer Materials}~\cite{Connolly2012}, the \textit{Handbook of Magnetic Materials}~\cite{Handbook} and the book {\it Magnetism and Magnetic Materials}.~\cite{Coey}. Nelson's database is then combined with $T_\mathrm{C}$ values from a dataset manually aggregated by Byland \textit{et al.}~\cite{Valentin-co}, which is mainly focused on, although not limited to, Co-containing compounds. Thus, this combined database is considered to be our ground truth, which amounts to 3,638 unique ferromagnetic compounds and their associated Curie temperatures.

Our results are also compared with a second database, the one obtained by combining the rules-based ChemDataExtractor scheme with a semi-supervised snowball algorithm~\cite{snowball}. At this point, it is important to remark that the two automatically extracted databases are based on a rather similar corpus of papers, namely those obtained with a keyword search from Crossref. This includes relatively recent articles and information is extracted only from the text. In contrast, the database of Nelson \textit{et al.} is largely based on data reported in tables and includes much historical information (containing results published as early as in the fifties). Despite the similarity in their respective corpora, however, the BERT-PSIE and ChemDataExtractor databases, of several thousand data points each, contain remarkably little overlap between each other, namely 694 compounds. The overlap between the automated and manual datasets is of similar size, namely 687 for BERT-PSIE vs.~manually-curated and 595 for ChemDataExtractor vs.~manually curated. Overall the three datasets (BERT-PSIE, ChemDataExtractor and the manually-curated one) share only 262 compounds. All comparisons performed in this section are done by taking the median Curie temperature value for compounds that contain multiple entries in each dataset.

\begin{figure*}[ht!]
    \centering
    \includegraphics[width=\linewidth]{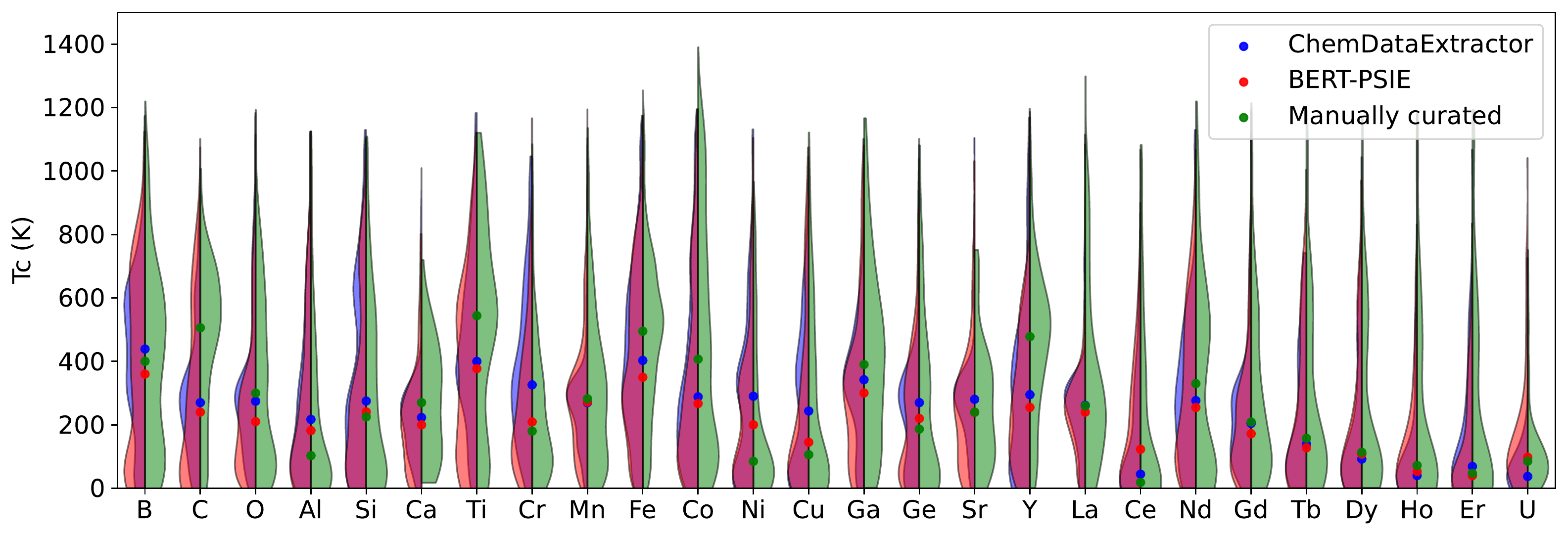}
    \caption[Violin plots comparing the $T_\mathrm{C}$ distribution of the compounds present in the various databases.]{Violin plots comparing the $T_\mathrm{C}$ distribution of the compounds containing specific elements in the dataset automatically generated with BERT-PSIE \textit{(red)} and ChemDataExtractor \textit{(blue)}, and in the manually-curated ground-truth \textit{(green)}. Only the most common elements appearing in the datasets are displayed here. The dots show 
    the median of each distribution.}
    \label{fig:TCDistribution}
\end{figure*}
Before going into the details of the comparison we observe that a significant source of error is associated with elemental compounds (e.g. Fe, Co, Ni, Gd, etc.), where the error is the variance of the extracted $T_\mathrm{C}$ compared with our ground-truth. This is due to the general difficulty of the NER in differentiating between an elemental compound and an element used as a dopant in an otherwise non-magnetic material (e.g. bulk Mn vs Mn-doped GaAs). As dopants can appear in a multitude of concentrations and in a large variety of hosts, erroneous assignments may result in a large spread in the distribution of the temperatures collated. With this exception, the distributions of Curie temperatures across the different databases are in very good agreement with each other, as can be seen in Fig.~\ref{fig:distro} (a). The agreement is particularly close between our automatically extracted dataset and the one constructed with ChemDataExtractor, but both present a peak in the distribution at around room temperature, which is absent from the manually curated one. There are two possible reasons behind this feature: either there is a bias in the most recent literature towards critical temperatures close to 300~K, or errors in the model aggregate $T_\mathrm{C}$ values close to ambient temperature. In support of the second hypothesis, it is worth recalling that mentions of room temperature feature heavily in sentences containing the target information, even if the room temperature is not the target temperature. An example of this situation is the sentence: `The magnetization curve at 300~K was obtained and the Curie temperature was determined by TGA under a magnetic field, yielding a Curie temperature of 1043~K for Fe.' Despite these differences, the still good similarity between the Curie-temperature distributions indicates that the relative abundance of high- and low-temperature ferromagnetic materials has been adequately captured by our automated extraction technique without the need for complex grammar-rule definitions.

Further understanding can be achieved by looking at the relative elemental abundance across the unique compounds present in a given database (the frequency with which a particular element appears in the database). This is shown in Fig.~\ref{fig:distro} (b), again for all three datasets. As expected the largest abundances are found for the magnetic transition metals, some of the rare earths and oxygen, a feature shared by all databases and corresponding to the actual elemental distribution among magnets. More interestingly, it appears that the automatically compiled databases overestimate the presence of Mn and O, and that of di- and tri-valent alkali metals (Ca, Ba, Sr and La). Such an overestimation with respect to the manually extracted dataset is significantly more pronounced for the ChemDataExtractor data than for the ones obtained with our workflow. It is likely that such a difference in distributions is mainly attributable to the data primary sources, which are different in the case of manually and automatically curated datasets. In particular, the most recent literature used in our extraction and in that performed by ChemDataExtractor contains many entries related to Ca-, Ba-, Sr- and La-containing perovskites (e.g. manganites). 

The influence of the primary data source on the final dataset is further confirmed by comparing the $T_{\mathrm{C}}$ distributions of compounds containing the 25 most common elements, which is presented in Fig.~\ref{fig:TCDistribution}. Generally, there is excellent agreement between the distributions of the two automatically extracted datasets, which contain entries extracted from similar sources. Then, BERT-PSIE generally captures a similar distribution to the manually extracted values, although there are evident discrepancies for certain elements. This may be an indication of the historical change in research focus between the sources used for the ground truth, compared with the sources for the automatically-extracted cases. 

\begin{figure*}[ht!]
    \centering
    \includegraphics[width=\linewidth]{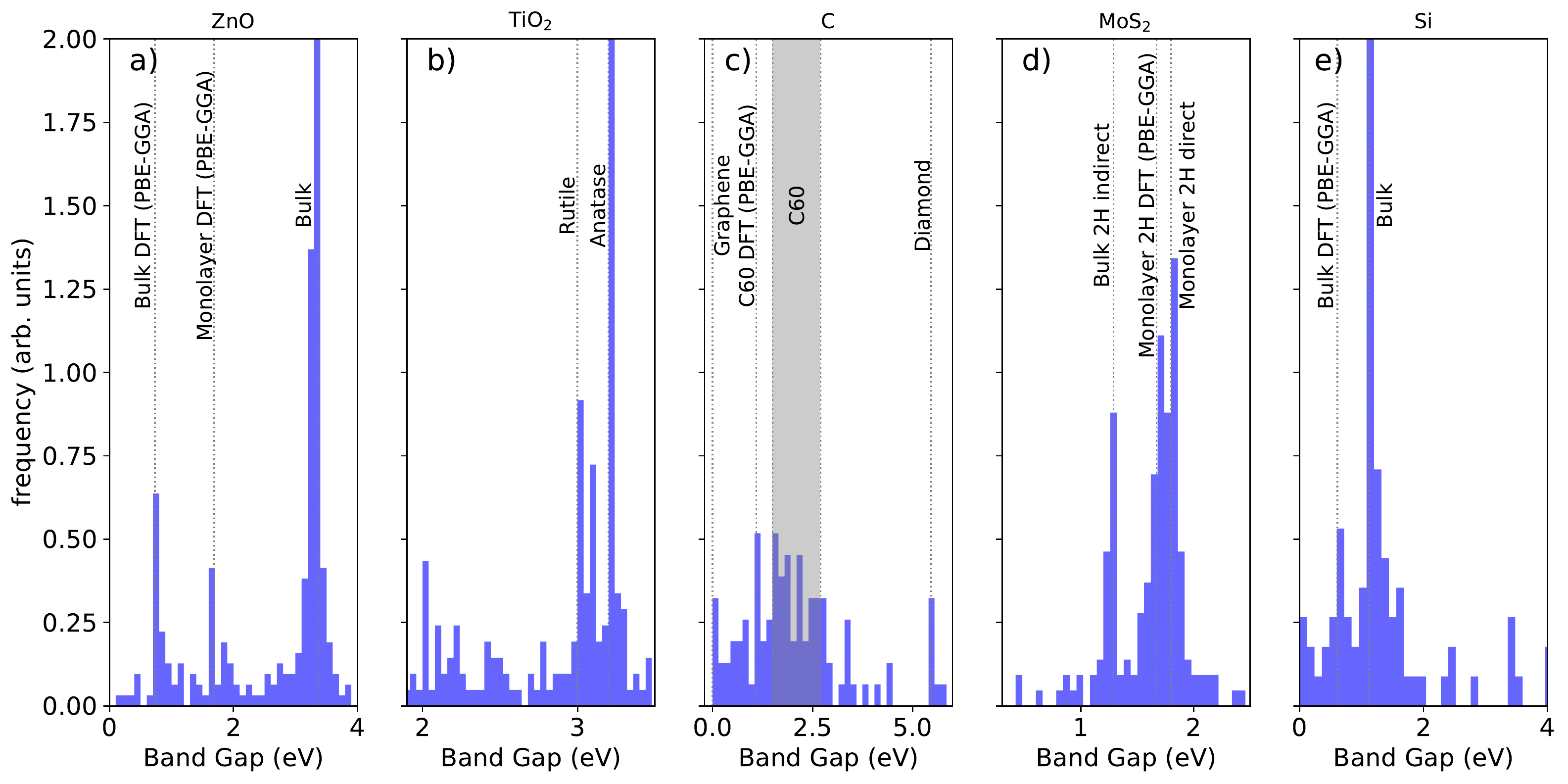}
    \caption[The distribution of band-gap values for the five most common chemical formulas found in the extracted database.]{The distribution of band-gap values for the five most common chemical formulas found in the  database of band-gaps generated by BERT-PSIE. The histograms report the relative abundance, while dashed lines indicate gap energies corresponding to specific experimental measurements or theoretical calculations.}
    \label{fig:top_5}
\end{figure*}

A similar study, with similar results, is performed on the distribution of the extracted band gaps, which can be found in Supplementary Figure 3. As for the case of Curie temperatures, we find strong similarities between our created database and the one obtained from ChemDataExtractor, but both present some level of disagreement with the manually curated one. 

We now discuss the origin of such disagreement by looking at the extracted band-gap distributions, see Fig.~\ref{fig:top_5}, of the five most common chemical formulas in the database, namely ZnO, TiO$_2$, C, MoS$_2$ and Si. The most interesting feature is that, while there is a spread of band-gap values for all five compounds, these are not uniformly distributed. In contrast, the band-gap densities seem to have a clear peak structure, with multiple high-frequency values appearing. This can be attributed to different means of obtaining the band gap of a material (experimental optical, experimental transport, theory, etc.), and to different polytypes, structures or dopant-varied compounds.

Going into more detail, consider first the case of ZnO in Fig.~\ref{fig:top_5} (a). In the distribution we find three clearly visible peaks, which are easily associated with the experimental bulk band-gap (3.37~eV \cite{znogga}), the density-functional-theory (DFT) calculated one for bulk ZnO (0.73 eV \cite{znogga} for PBE-GGA) and the DFT-calculated one for monolayer ZnO (1.69 eV \cite{znoggamono} again PBE-GGA). A similar situation is encountered for Si, presented in Fig.~\ref{fig:top_5} (e), where the two main peaks are attributed to the experimental bulk indirect gap (1.1 eV \cite{silicon}) and the one returned by DFT simulations (0.61 eV \cite{CURTAROLO2012227}, PBE-GGA). A different situation is then encountered for TiO$_2$ in Fig.~\ref{fig:top_5} (b), where the two main dominant peaks both correspond to experimental gaps, but they are for two different polymorphs, namely anatase (3.2~eV) and rutile (3.0~eV) \cite{tio2}.

Finally, $\mathrm{MoS_2}$ in Fig.~\ref{fig:top_5} (d), and carbon in Fig.~\ref{fig:top_5} (c), display more complexity. In the first case, we find three dominant peaks. In fact, together with the experimental, bulk, indirect band-gap of 1.29~eV \cite{mos2bulk}, many mentions in literature concern the experimental band-gap of the monolayer form of MoS$_2$ (1.8~eV \cite{mos2}) and the DFT estimate of the same (1.67~eV \cite{mos2gga}, PBE-GGA). Carbon, in contrast, deserves attention on its own, since a large variety of polymorphs are possible. In fact, the distribution shows a clear peak for semimetal graphene  \cite{doi:10.1126/science.1102896} and one for the bulk diamond structure (5.47~eV \cite{diamond}). Then, there is a uniformly distributed region, which is characterised by band-gap values associated with carbon buckminsterfullerenes, C60. This extends over the 1.5-2.7~eV range, and has a clear peak at the DFT value of 1.09~eV (PBE-GGA)~\cite{c60gga}.

\subsection{Database Quality for Downstream Tasks: Curie Temperature}\label{sec:DataQuality}
\begin{figure}[h]
    \centering
    \includegraphics[width=0.9\linewidth]{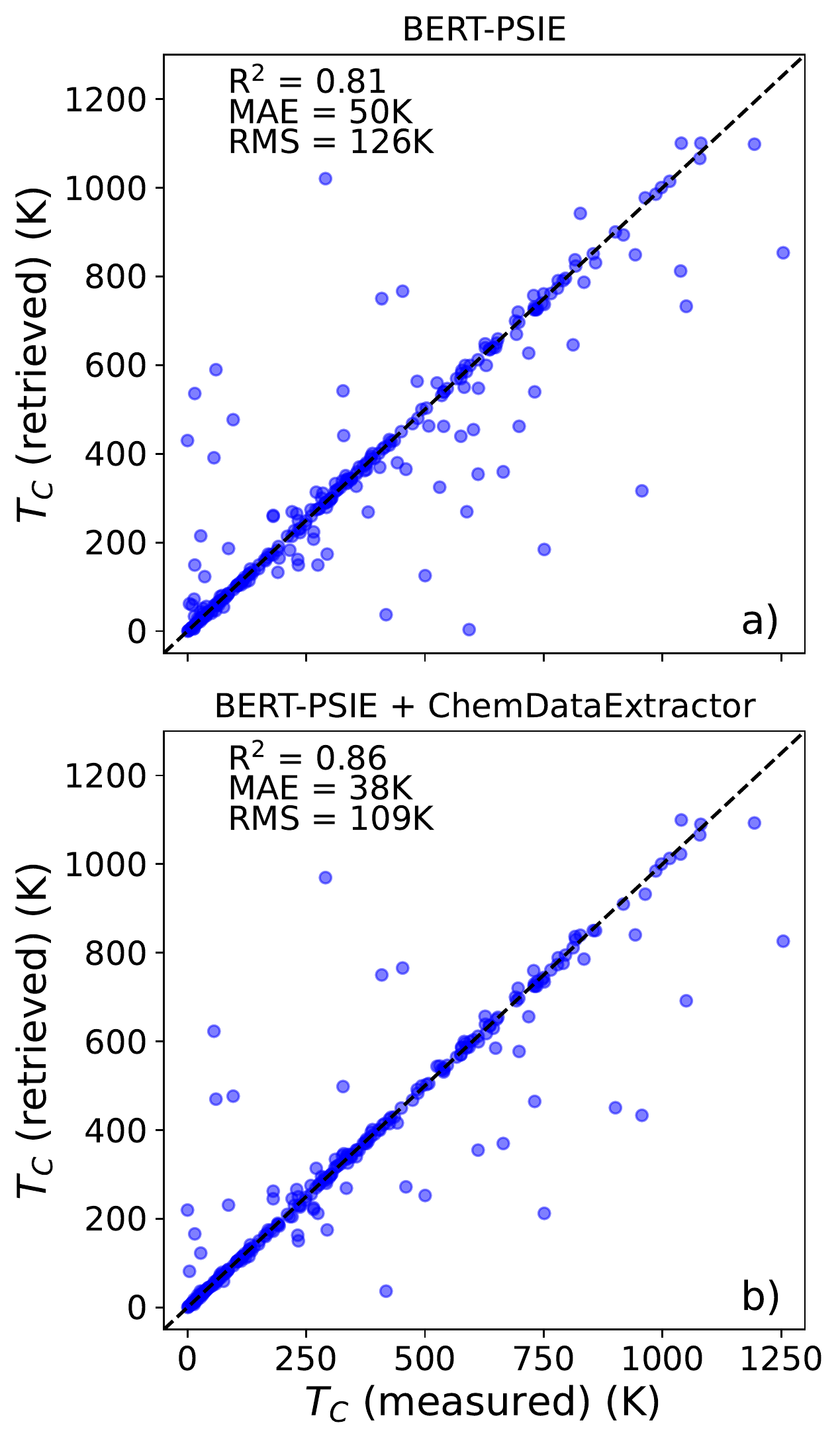}
    \caption[Comparison between the $T_\mathrm{C}$ queried from the manual and BERT-PSIE dataset.]{\textbf{(a)} Comparison between the $T_\mathrm{C}$ queried in the dataset automatically generated by BERT-PSIE and the values contained in the manually-curated dataset. The comparison is performed over the 262 compounds that are shared by all datasets examined in this work. The median value is returned whenever multiple $T_\mathrm{C}$ values are collected for a given compound. \textbf{(b)} The same comparison is performed on the dataset resulting by combining the one generated by BERT-PSIE and the one generated by ChemDataExtractor.}
    \label{fig:query}
\end{figure}

The quality of a material-property database can be quantified in terms of its usefulness for material design. To assess this, we introduce two tests that simulate real-world usage of these databases of material properties. The first is a `query test' that evaluates the quality of the retrieved data. The second is a `suitability for machine learning test' that gauges the data's capability to support the training of machine learning models. Once again, we start the discussion from the Curie temperatures, by firstly assessing that the returned value to a query related to a compound present in the database is reliable. To achieve this goal, we have designed a `query test' comparing the Curie temperatures automatically extracted with the one present in our reference manually extracted dataset. In order to make the comparison between our database and the ChemDataExtractor-generated one not dependent on the particular class of compounds extracted, we only compare entries that are shared by all the datasets (262 compounds). The query test results for BERT-PSIE are reported in Fig.~\ref{fig:query}, while the performance metrics for the different datasets are summarised in the left-hand side of Table~\ref{table:data_comp}.
\begin{table*}[t]
\caption{Performance comparison between the different automatically generated $T_{\mathrm{C}}$ datasets against the manually-curated one from    Ref.~[\onlinecite{James_tc,Valentin-co}]. The left-hand side of the table refers to the query test, while the right-hand side refers to the RF $T_\mathrm{C}$ predictor. Together with the BERT-PSIE and ChemDataExtractor databases we also consider different BERT-assembled datasets obtained by using different relation-classification strategies (see details in the text). The query benchmark is done over the 262 compounds that are shared by all the datasets, while the RF predictions are done over 2,623 compounds that are not present in any of the automatically collated datasets. Values for the best-performing datasets are in bold. Additionally, the last line of this table shows the performance on these tests for the combined databases generated by BERT-PSIE and ChemDataExtractor, leading to the best results.}\label{table:data_comp}
\centering
\begin{tabular}{c | c | c | c | c | c | c | c }
  \hline
  & $\;\;$\# Entries$\;\;$ & \multicolumn{3}{c|}{Query} & \multicolumn{3}{c}{RF Predictions} \\ 
  \hline
  & & $\;\;\;\;\mathrm{R^2}\;\;\;\;$  & MAE (K) & RMSE (K) & $\;\;\;\;\mathrm{R^2}\;\;\;\;$ & MAE (K) & RMSE (K) \\
  \hline
  \hline
  ChemDataExtractor & 4,289 & 0.78  & \textbf{48} & 137 & 0.65 & \textbf{123} & 176 \\
  \hline
  \hline
  \multicolumn{8}{c}{This work}\\
  \hline
  Single Mentions & 1,858 & 0.77  & 51 & 139 & \textbf{0.66} & 128 & \textbf{174} \\
  \hline
  Order of Appearance & 2,682 & 0.77  & 51 & 141 & 0.65 & 126 & 176 \\
  \hline
  All Combinations & 4,308 & \textbf{0.81}  & 52 & 127 & 0.61 & 134 & 184 \\
  \hline
  BERT-PSIE & 3,518 & \textbf{0.81}  & 50 & \textbf{126} & 0.65 & 126 & \textbf{174} \\
  \hline
  \hline
  BERT-PSIE + ChemDataExtractor & 7,052 & 0.86  & 38 & 109 & 0.69 & 118 & 165 \\
  \hline
 \end{tabular}
 
\end{table*}

As discussed in the previous section, the most challenging step in our extraction workflow is the relation-classification step. In order to evaluate the performance of our model for relation classification a variety of additional extraction strategies have been attempted and compared against the ground-truth dataset. The first of these strategies involves as before taking only the `Single Mentions' results extracted from sentences containing only a single mention of a compound and a single mention of a Curie temperature value. In this case, we assume that the two entity mentions are related to each other, thus removing completely the need for any relation-assignment step (`Single Mentions' in Table~\ref{table:data_comp}). The second strategy imposes a rule that associates compound/value pairs based on the order in which they appeared in the text (`Order of Appearance' in Table~\ref{table:data_comp}). Finally, we have taken every possible combination of compound/value pairs, in order to compare our results with random associations (`All Combinations' in Table~\ref{table:data_comp}). This choice corresponds to a relation-classifier model that always outputs a positive classification. Table~\ref{table:data_comp} is complemented by results obtained with our constructed BERT classifier (`BERT-PSIE'), with the data extracted by ChemDataExtractor and by aggregating these last two datasets (`ChemDataExtractor + BERT-PSIE').

As can be seen from Table~\ref{table:data_comp}, all of the datasets compiled with our rule-free pipeline have metrics comparable to those of ChemDataExtractor, and the one constructed with BERT-PSIE appears to be the best performing on almost all the query-test metrics. In particular, BERT-PSIE returns the best $R^2$ coefficient of 0.81 and root mean squared error (RMSE) of 126~K. Interestingly, BERT-PSIE gives us a mean absolute error (MAE) slightly larger than that obtained with ChemDataExtractor. This suggests that BERT-PSIE achieves an accuracy on par with ChemDataExtractor (the two produce datasets equally similar to the manually curated one), but it is slightly less prone to display large outliers. Note, however, that the notion of an outlier and its relevance to the overall performance of a method need to be taken with some caution in this query test. In fact, if the $T_\mathrm{C}$ of a compound is erroneously extracted (the ML model extracts the wrong temperature), there is no particular advantage of having the wrong $T_\mathrm{C}$ close to the real one. This point can be better understood by looking at the parity plot in Fig.~\ref{fig:query}, where entries are either on the parity line (exact extraction) or away from it without any particular correlation with the actual $T_\mathrm{C}$ value (erroneous extraction).

In any case, the fact that BERT-PSIE performs better than any other BERT-based models using different relation-classifier strategies demonstrates that the inclusion of a context-aware mean of extracting compound-value pairs from literature is advantageous. Unfortunately, this is not a massive improvement, since the metrics are rather close to those obtained by considering randomised combinations of all possible compound-value pairs (`All Combinations'). Indeed, more sophisticated methods to establish the correct compound-property associations will help in producing a better-automated dataset. 

Our second test probes the ability of a given data-extraction strategy to create datasets of sufficient quality to enable the construction of predictive ML models. In practice, we want to establish whether the data extracted can be the platform for models that predict the $T_\mathrm{C}$ of unseen compounds. In particular, we target compositional models, which are ML algorithms using information directly accessible from the chemical composition of a given compound as features. For the case of the Curie temperature associated with the ferromagnetic transition, in fact, it has been shown that compositional models can achieve good performance if trained on manually-curated data~\cite{James_tc} (note that the model mentioned does not describe other magnetic phases, e.g antiferromagnetic structures). In order to test the ability of a dataset to function as a reliable data platform for ML models, we train on each automatically generated dataset a random forest (RF) model that takes as input compositional features, as done in Reference [\onlinecite{James_tc}] and [\onlinecite{Wolverton}]. We have chosen the same input features for all the RF models trained, since in all the cases considered we have not observed any improvement when adding more features. We then compare the predictions on a set of compounds that are not present in the training set with the values extracted manually from the literature. For this test, we consider predictions over 2,623 compounds for which we have a manually extracted $T_\mathrm{C}$ that do not appear in any of the datasets automatically extracted. With this choice, both our tests are performed on the same compounds for all the datasets considered. For compounds with multiple values of extracted $T_\mathrm{C}$, the median value of the collated results is taken as the associated Curie temperature, according to the procedure introduced in Ref.~[\onlinecite{James_tc}]. We have also tested other summary statistics, such as the mean and the mode, without finding any significant difference in the results. 

Again the results of our test are reported in Table~\ref{table:data_comp} (right-hand side), where one can clearly see that BERT-based extraction workflows perform rather similarly to the established rule-based method. In particular, the full workflow, BERT-PSIE, has an $R^2$ identical to that obtained by ChemDataExtractor, with a better RMSE but worse MAE, we observe for this second test a result similar to that found for the query test. Most interestingly, we find that the inclusion in the database of entries extracted in conjunction with the relations-classification step does not improve the performance of the predictor. In fact, using single mentions only returns us the better $R^2$ value of 0.66 and an RMSE of 174~K, while BERT-PSIE gives us a slightly degraded $R^2$ at 0.65 and an identical RMSE, although it slightly improves the MAE (by about 2~K). This is possibly due to the fact that the inclusion of the entries from multiple mentions inherently adds noise to the database. Thus, despite the fact that the model can be trained over a much larger dataset, no significant improvement is detected. 
\begin{figure}[h!]
    \centering
    \includegraphics[width=0.9\linewidth]{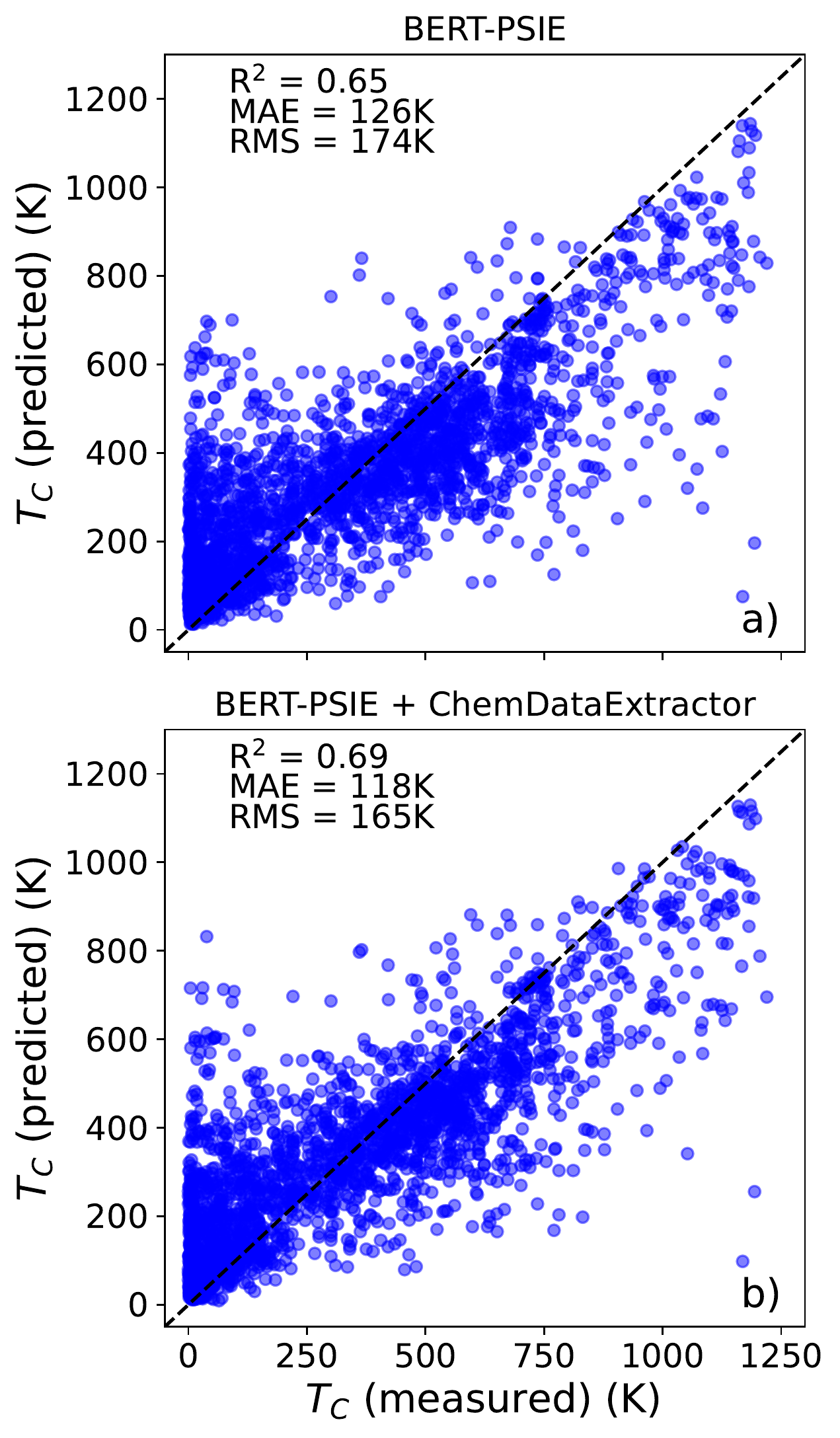}
    \caption[Parity plot for the predicted $T_\mathrm{C}$ from BERT-PSIE against the manually-extracted $T_\mathrm{C}$.]{Parity plot (predicted $T_\mathrm{C}$ vs manually extracted $T_\mathrm{C}$) for the best RF compositional model constructed \textbf{(a)} on the BERT-PSIE dataset and \textbf{(b)} on the combined BERT-PSIE and ChemDataExtractor dataset. The test set consists of the 2,623 compounds that are not present in any of the automatically generated datasets considered in this work, but for which we have a $T_\mathrm{C}$ manually extracted from the scientific literature.}
    \label{fig:RF}
\end{figure}

The parity plot of our best RF model trained on the full BERT-PSIE dataset is presented in Fig.~\ref{fig:RF}. In general, the $T_\mathrm{C}$ trends are captured, but it is also clear that the model is significantly inferior to that presented in Ref.~[\onlinecite{James_tc}], which reports a MAE of 57~K roughly a factor of two smaller than the 126~K obtained for a random forest model trained on the data extracted with BERT-PSIE.

Although this can be partially attributed to noise in the data, for instance to the likely presence in the BERT-PSIE dataset of critical temperature associated with antiferromagnets, one has also to consider that the data used in Ref.~[\onlinecite{James_tc}] were highly curated even after the collection. For example, additional data on paramagnets was included to improve the low-temperature part of the distribution, while data corresponding to different concentrations of metallic alloys was selectively excluded, to better balance the chemical distribution. All these post-processing steps were not performed here, since our task is simply that of assessing the quality of the automatically compiled dataset. In fact, one expects that automatically constructed datasets can reach sizes large enough for such post-processing steps to not be necessary.

Given the fact that the overlap between our BERT-PSIE dataset and the one generated by ChemDataExtractor consists of only 694 compounds, we have constructed an additional dataset resulting from the combination of the two. This combined database contains 7,052 distinct entries and performs best on all the metrics evaluated in each test as seen in the last line of Table~\ref{table:data_comp} and in Fig. \ref{fig:query} (b) and Fig. \ref{fig:RF} (b). The improvement in performance is likely due to the much larger size of the dataset (approximately double the original two) and the corresponding reduction of the noise present in the median values.  When tested against RF models, the much larger number of compounds allows for a better sampling of the chemical space, resulting in more accurate predictions. As it stands, this combined dataset represents the best database available for ferromagnetic $T_\mathrm{C}$, automatically extracted from scientific literature according to the test designed here. The implication of this is that the quality of automatically extracted databases improves markedly with an increase in the number of disparate sources used in the extraction. There is also an argument to be made that a combination of rules-based and rule-free methods may be the best-performing strategy for automated extraction as seen in Table~\ref{table:direct_comp}.

A similar study was conducted with respect to the extractions performed by ChemDataExtractor over the same sentences deemed relevant by the BERT-PSIE sentence classifier module. The results can be found in the Supplementary Discussion and reinforce the similarity in performance of the two approaches.



%
\begin{figure}[h]
    \centering
    \includegraphics[width=\linewidth]{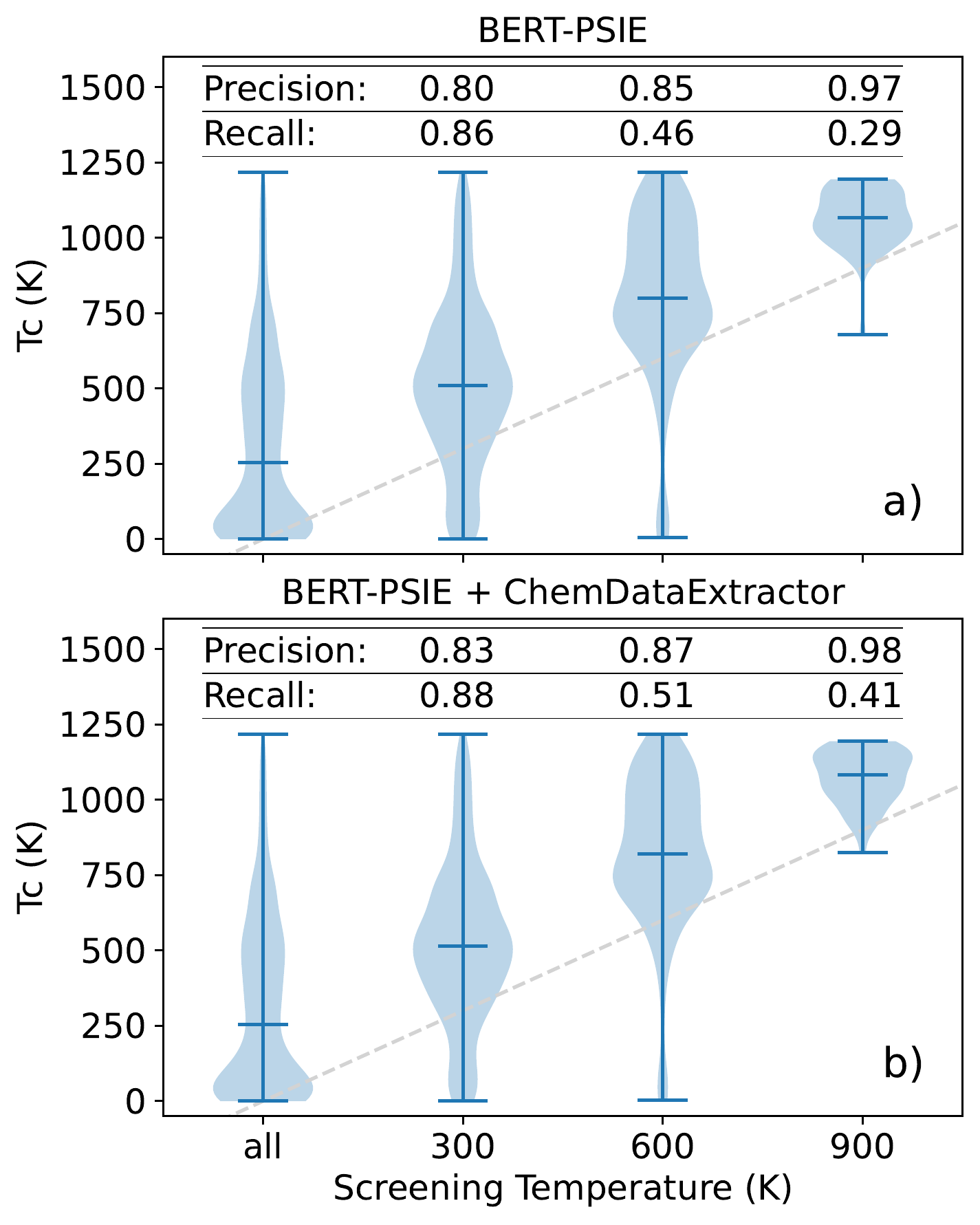}
    \caption[Violin plots showing the $T_\mathrm{C}$ distributions of the compounds screened for different temperature brackets.]{\textbf{(a)} Violin plots showing the $T_\mathrm{C}$ distributions of the compounds screened using a RF model trained on the BERT-PSIE data and compared with the manually extracted values. The dashed line is the parity line highlighting how the median of the screened distribution increases as the screening threshold increases. Despite a low recall, the precision is high enough to select compounds likely to have a $T_\mathrm{C}$ higher than a given threshold. The screening is done on compounds not present in the training set of the RF. \textbf{(b)} The same test is performed by training a RF model on the combination of the BERT-PSIE and ChemDataExtractor datasets.}
    \label{fig:violin}
\end{figure}
To conclude, as a final evaluation of the usefulness of the extracted database, we test the ability of the RF model trained on the BERT-PSIE dataset to screen unseen compounds with respect to a certain $T_\mathrm{C}$ threshold. This test attempts to simulate a common use case for such ML models. Note that typical magnets employed as part of some room-temperature technology (e.g. data storage, electrical motors) need to have a $T_\mathrm{C}$ of the order of 600~K so that classifying magnets according to such a threshold is of significant technological relevance. We used the RF model trained on the automatically generated dataset to predict if magnets have a critical temperature exceeding 300~K, 600~K and 900~K, respectively. The test set for this predictor is constructed from compounds that are present in our manually-curated dataset, but not in the one generated by NLP. The results of this screening, compared with the distribution of the true $T_\mathrm{C}$ of these compounds can be seen in Fig.~\ref{fig:violin}. The shaded blue area of Fig.~\ref{fig:violin} represents the distribution of values predicted to have a $T_\mathrm{C}$ greater than the dashed line, representing the screening temperatures of 300~K, 600~K and 900~K respectively. While the recall of this screening is quite low, the high precision biases the initial distribution into sets with higher and higher $T_\mathrm{C}$, thus demonstrating the usefulness of the extracted database in screening for compounds with $T_\mathrm{C}$ above a desired threshold. The low recall means that certain compounds with $T_\mathrm{C}$ above the desired temperature will not be predicted to be in the set of compounds above the temperature, however, the compounds predicted to be in this set can be trusted with high accuracy to have a $T_\mathrm{C}$ above the desired threshold. Training a model using the combined datasets from BERT-PSIE and ChemDataExtractor results in a higher screening recall. This provides an example of how the utility of these automated databases can be improved both by expanding the corpus size used for extraction and by introducing new extraction techniques.

\subsection{Database Quality for Downstream Tasks: Band-Gap}\label{sec:DataQualityBG}

\begin{table*}[t]
\caption{Performance comparison between the different automatically generated band-gaps datasets against the manually-curated one from Ref.~[\onlinecite{BandGapData}]. The left-hand side of the table refers to the query test, while the right-hand side refers to the RF band-gap predictor. Together with the databases constructed using BERT-PSIE and ChemDataExtractor, we also consider different BERT-assembled datasets obtained by using different relation-classification strategies (see details in the text). The query benchmark is done over the 231 compounds that are shared by all the datasets, while the RF predictions are done over 2046 compounds that are not present in any of the automatically collated datasets. Values for the best-performing datasets are in bold.}\label{table:data_gap}
\centering
\begin{tabular}{c | c | c | c | c | c | c | c }
  \hline
  & $\;\;$\# Entries$\;\;$ & \multicolumn{3}{c|}{Query} & \multicolumn{3}{c}{RF Predictions} \\ 
  \hline
  & & $\;\;\;\;\mathrm{R^2}\;\;\;\;$  & MAE (eV) & RMSE (eV) & $\;\;\;\;\mathrm{R^2}\;\;\;\;$ & MAE (eV) & RMSE (eV) \\
  \hline
  \hline
  ChemDataExtractor & 2185 & 0.54  & 0.78 & 1.34 & 0.59 & $\mathbf{0.62}$ & 0.87 \\
  \hline
  \hline
  \multicolumn{8}{c}{This work}\\
  \hline
  Single Mentions & 1,246 & 0.65  & 0.67 & 1.17 & 0.61 & $\mathbf{0.62}$ & 0.85 \\
  \hline
  Order of Appearance & 1819 & $\mathbf{0.67}$  & $\mathbf{0.64}$ & $\mathbf{1.13}$ & $\mathbf{0.62}$ & 0.63 & $\mathbf{0.84}$ \\
  \hline
  All Combinations & 2581 & 0.63  & 0.71 & 1.21 & 0.60 & 0.63 & 0.86 \\
  \hline
  BERT-PSIE & 2021 & 0.64 & 0.67 & 1.19 & 0.61 & $\mathbf{0.62}$ & 0.85 \\
  \hline
\end{tabular}
\end{table*}

To further validate the performance of BERT-PSIE, a similar study is performed, now with the target being an aggregated dataset of compounds and their associated band-gap. For the manually-curated test set in this instance, a database of band-gaps from reference~\cite{BandGapData} is utilised and compared with our results and with the results of the hybrid ChemDataExtractor model from Dong {\it et al.}~\cite{Dong2022} when ran on the same literary corpus. The original dataset from this paper is not used as a direct comparison between the models as the workflow implemented in Ref.~\cite{Dong2022} also separately processes tables, which are not considered by BERT-PSIE. The results of the comparison between the two methods on the same corpus can be seen in Table~\ref{table:data_gap}. In this case, the BERT-PSIE pipeline outperforms or equals the hybrid ChemDataExtractor method by every metric, while extracting a very similar number of unique compound-band-gap relationships. Interestingly, for sentences containing multiple mentions the best strategy to sort out relations seems to be the order of appearance, which outperforms all other methods. This is in contrast with the degradation in performance reported for the case of Curie temperature, Table~\ref{table:data_comp}, pointing to an intrinsic difference in the way these two quantities are reported in natural language. It then appears that reporting the band gap is far more procedural than reporting the Curie temperature, thus the use of a more sophisticated method of establishing the correct associations between compounds and properties introduces a source of noise. This result is clearly property dependent, but the difference in performance is marginal.

Finally, the parity plot for the query test and that for a RF model for band-gap predictions are shown in Supplementary Figure 4. In the first case, the results are similar to those found for the Curie temperature, although it seems that a more diffuse distribution of band gaps is now observed. This is associated with the ambiguity in the various band-gap definitions noted before (e.g. the case of C), an ambiguity that is less relevant for the $T_\mathrm{C}$. The RF model, instead, appears to have a slightly inferior $R^2$ than that constructed for the $T_\mathrm{C}$ but benchmarks similarly with models that can be constructed on manually-curated data. In fact, we obtain a MAE of 0.62~eV, against the value reported on MatBench~\cite{matbench} of 0.33~eV, for the best-performing model trained on the same dataset.

\section{Discussion}

We have proposed a workflow to automatically extract structured data from unstructured scientific literature. 
This has minimal need for an extensive implementation effort and little or no requirement for familiarity with 
complex grammar-rule definitions and natural language processing. We have then shown some possible use cases, 
demonstrating the ability to generate a database of ferromagnetic Curie temperatures and electronic band 
gaps comparable to the one generated using ChemDataExtractor, the state-of-the-art rule-based method for 
data mining from the scientific literature. This work opens the door for rapid and easy access to experimental-property 
databases for materials informatics applications.

Crucially, we have carefully benchmarked the constructed databases against manually curated 
reference ones, through extensive query tests. These have allowed us to critically assess the benefit of certain design 
choices in our workflow, such as the relation-extraction step. Most importantly, we have been able to understand where 
improvements can be made and whether these are general or specific to the physical property extracted.

Finally, we have tested whether our automatically extracted databases of compounds and their properties can be 
used as a platform for constructing machine-learning models, namely whether the database quality is sufficiently 
high for integration in a material-discovery workflow. We have found that the mean absolute error of chemical-informed
random-forest models, constructed over the automatically extracted database, is always larger than that achieved with 
manually curated ones, roughly by a factor of two. Our best results have been obtained for a Curie-temperature database 
combining data from rule-free (BERT-PSIE) and rule-based (ChemDataExtractor) methods, owing to the larger data 
volume and enhanced diversity. Notably, no manual curation was performed on the automatically extracted datasets, a 
fact that may be responsible for the differences in performance. In contrast, we have shown that our BERT-PSIE dataset 
is sufficient to construct machine-learning classifiers able to identify high-$T_{\mathrm{C}}$ magnets with high accuracy. 
This fact, together with the possibility to expand the dataset with the minimum effort offered by a rule-free strategy, 
suggests that natural-language-processing information retrieval can become an important asset in any material-discovery 
pipeline.

\section{Methods}
In this section, we outline the NLP workflow implemented in BERT-PSIE for the automated extraction of structured data from scientific literature. This is based on the concatenation of different BERT models, fine-tuned to perform the specific tasks necessary for the text-mining pipeline. Our designed workflow can be adapted to the extraction of any binary-related information and can potentially be extended to more complex form relations. In this work, we focus on mining the $T_\mathrm{C}$ of ferromagnets and the electronic band-gap of semiconductors/insulators because of the availability of manually-curated databases, which can be used as ground truth in assessing the performance. It is certainly true that the extraction of inter-related properties (for instance, a material's property may depend on the experimental temperature) can be more complex, but these are difficult to benchmark because of the lack of manually curated datasets. In such a case, the extraction should provide the compound and all the inter-related quantities, simultaneously. The fact that BERT-PSIE works at the sentence level may pose a limit to this, since the different information may be contained in different sentences. Strategies to overcome such limitations certainly deserve future work.

%
\begin{figure*}
    \centering
    \includegraphics[width=\linewidth]{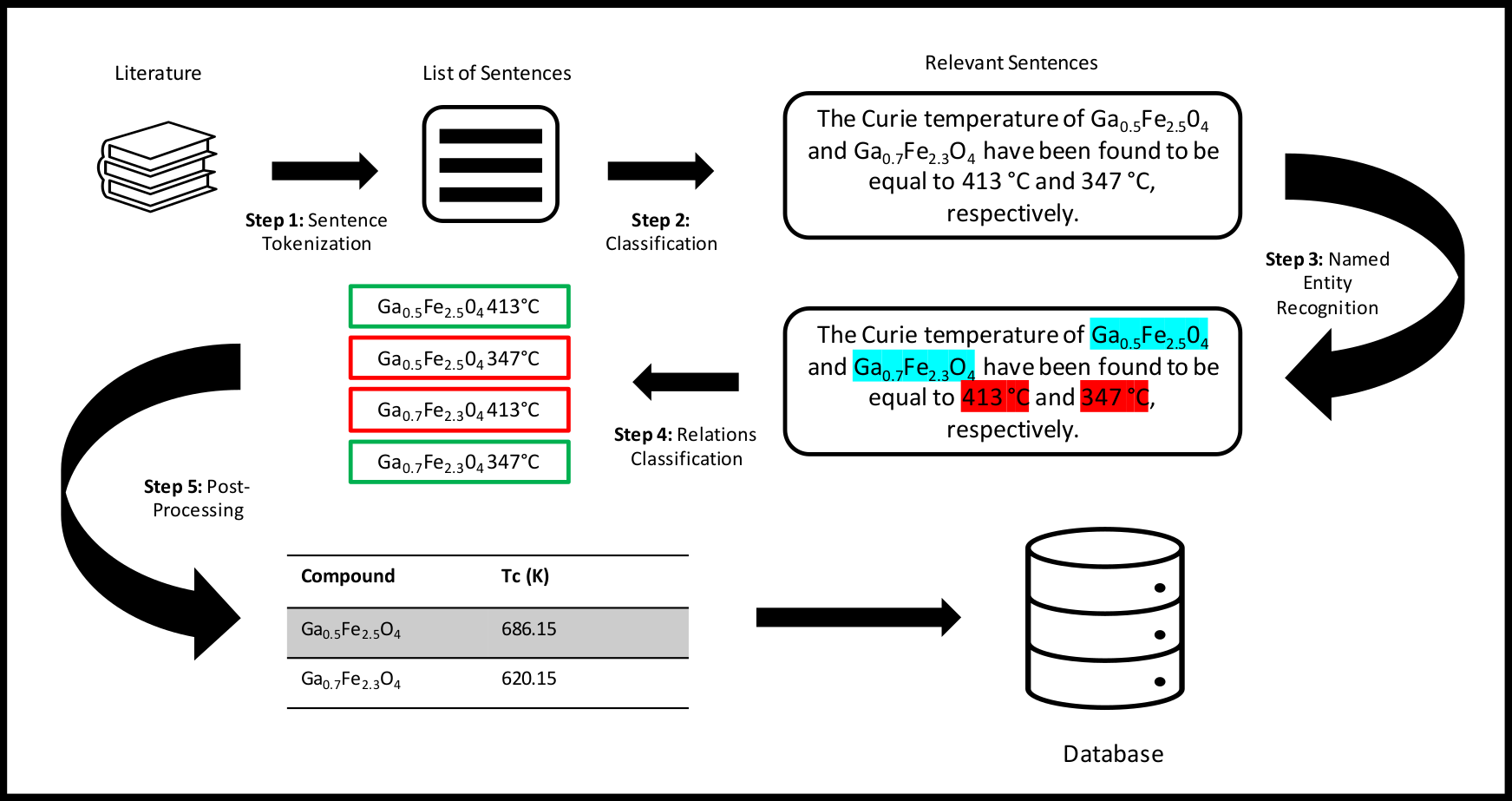}
    \caption[The BERT-PSIE workflow for the automated extraction of scientific information from unstructured literature.]{Schematic diagram of the BERT-PSIE pipeline for the automated extraction of compound-property pairs from scientific literature. The workflow relies on the combination of BERT models fine-tuned for different downstream tasks such as sentence classification, named entity recognition and relation classification. Here we use the Curie temperature as an example. See text for more details.}
    \label{fig:pipeline}
\end{figure*}

\subsection{BERT-PSIE Structure}
The structure of our entire workflow can be appreciated by looking at the scheme in Fig.~\ref{fig:pipeline}. Papers are downloaded from the web by using a keyword-based search with the Crossref REST API (api.crossref.org). A classifier identifies the relevant sentences from the downloaded corpus and a Named Entity Recognition (NER) module extracts material-property relations from sentences containing single unambiguous relations. Note that sentences containing a single entity only, either the compound or the property, are discarded. Then, a second module performs relation classification for sentences where multiple mentions of compounds and/or materials' properties are present. The material-property relations extracted then form the database. 

All of the BERT models used in this work have been fine-tuned on a single Nvidia A100 GPU accessed with Google Colab, with each model requiring less than 30 minutes for fine-tuning. In particular, we use an early-stopping strategy based on the validation-set loss (more details are provided in the Supplementary Methods). As it stands, the time bottleneck in adapting the workflow to a new task remains related to the manual labelling of the text necessary for fine-tuning. All in all, this takes about one week. We now describe in more detail the algorithms and the training strategy used for the various models.

\subsection{Literature Extraction}\label{sec:lit_extract}
The Crossref REST API is used to execute a keyword search over all literature published by Elsevier. This yields metadata filtered to ensure that the full-text version of the paper is available for the purpose of text data mining. The metadata includes both abstracts and download links to the full-text papers. In the case of the Curie temperature extraction, the strategy used to build the training set required for the fine-tuning of the different BERT models starts from the collection and the manual labelling of 800 abstracts containing the term `Curie temperature' from this initial search. We run the Natural Language Toolkit~\cite{nltk} (NLTK) sentence tokenizer on these abstracts and label the tokenized sentences, which reference a Curie temperature, as relevant and the ones that do not as irrelevant (Step 1 in Fig.~\ref{fig:pipeline}). This step yielded a database of approximately 5000 sentences of which 189 are labelled relevant. The labelled dataset is used to fine-tune a BERT classifier model to find relevant sentences (i.e. sentences likely to contain a mention of the Curie temperature). The classifier extracts relevant sentences from the corpus of the papers. We then manually labelled 200 relevant sentences extracted from the corpus of the papers whose abstract was used in the previous step. These extracted sentences are manually labelled as described in Step 3 of Fig.~\ref{fig:pipeline} and are then combined with the labelled abstracts. This combined corpus is utilised to fine-tune a BERT model for Named Entity Recognition (NER-BERT).

In the case of the electronic band gap, a slightly different strategy is developed for aggregating the necessary 
training data. The arXiv metadata is downloaded from the Kaggle dataset \cite{arXivData} and an initial corpus of 1,000 abstracts for annotation is constructed by searching the text of the corpus of abstracts for the terms `band gap', `bandgap' or `band-gap'. Contained in these 1,000 abstracts were 171 sentences that contained band-gap values and thus were considered relevant. A sample of 501 sentences that contained no mention of band gap, and were thus considered irrelevant, were added to the relevant ones. This enabled the creation of a classifier training set of 672 sentences.

The corpus used for the automatic extraction procedure of the Curie temperature (band-gap) is obtained by executing a keyword search using the Crossref API for instances of the term `magnetic' (`electronic'). This yields a database of full-text URLs of papers likely to contain a mention of a Curie temperature (band-gap) value. The papers are then automatically downloaded and parsed into a list of sentences. A corpus of relevant PDF documents is also converted into plain text using PDFminer~\cite{pdfminer} and is similarly parsed into sentences, which are concatenated to the same list. A list of candidate sentences likely to contain the desired material property information is then extracted using the BERT classifier from these corpora.

\subsection{Relation extraction}
The final step of our workflow for the automatic extraction of data consists of the identification of mentions of chemical compounds and the associated property (Curie temperature or band-gap) in all the sentences classified to be likely to contain such information. This task is performed by the NER-BERT model (Step 3), which is described here for the Curie temperature (the same applies to the band gap). For the sentences predicted by the NER-BERT model to have a single mention of a chemical compound and a single mention of Curie temperature, we assume that the two quantities are related and we add them to the database (Step 5). If a sentence contains multiple mentions of chemical compounds and/or several Curie temperatures, the compound-temperature association will become ambiguous. This ambiguity is not uncommon in scientific literature, where one can find sentences like `the Curie temperature of Fe and Co are 1043~K and 1394~K, respectively'. Although easy to resolve for a human reader, semantic ambiguity becomes a problem for NLP. Here, we treat the problem as a relation classification task. In practice, following the approach of Soares {\it et al.}~\cite{soares1906matching}, we fine-tune a BERT architecture to classify whether a pair of entities in a sentence is related by the `has a $T_\mathrm{C}$ of' relation.

The dataset needed for the fine-tuning is generated by sampling 100 sentences from among those predicted to contain multiple mentions by the NER-BERT model. For each sentence, all the possible pairs of compound-$T_\mathrm{C}$ mentions are considered one by one, and entity markers are added at the beginning and at the end of each entity mention. For example, from a sentence containing two chemical compounds and two Curie temperatures, we generate four sentences, in which a different pair of entities is surrounded by entity markers. We use the markers $[E1_\mathrm{start}]$, $[E1_\mathrm{end}]$ to identify the compound mentions considered and $[E2_\mathrm{start}]$ and $[E2_\mathrm{end}]$ to identify the Curie temperature mention. Thus, by taking as an example the sentence, `The Curie temperature of Ga$_{0.5}$Fe$_{2.5}$O$_4$ and Ga$_{0.7}$Fe$_{2.3}$O$_4$ have been found to be equal to $\mathrm{413~^{\circ}C}$ and $\mathrm{347~^{\circ}C}$, respectively' (see Fig.~\ref{fig:pipeline}), we construct the following four associations: 1) Ga$_{0.5}$Fe$_{2.5}$O$_4$ and $\mathrm{413~^{\circ}C}$, 2) Ga$_{0.5}$Fe$_{2.5}$O$_4$ and $\mathrm{347~^{\circ}C}$, 3) Ga$_{0.7}$Fe$_{2.3}$O$_4$ and $\mathrm{413~^{\circ}C}$, 4) Ga$_{0.7}$Fe$_{2.3}$O$_4$and $\mathrm{347~^{\circ}C}$. Then, the sentences with the marked pairs generated are manually labelled for a binary classification task, where we will deem a sentence positive if the relation `has a $T_\mathrm{C}$ of' is present between the two marked entity mentions, and negative if such a relation is not present. The resulting training set, after being balanced with respect to the mentions in each class, consists of 200 sentences each one containing a different pair of marked entities. This collection of BERT models trained for different downstream tasks creates a rule-free pipeline for the automatic extraction of data from text.

\section*{Data Availability}

The test abstracts for the direct comparison with rule-based methods together with the datasets automatically extracted in the main and test extractions are available at \url{https://github.com/StefanoSanvitoGroup/BERT-PSIE-TC}.

\section*{Code Availability}

All the BERT-PSIE code necessary to run the automatic extraction of Curie temperatures and band gaps is available at \url{https://github.com/StefanoSanvitoGroup/BERT-PSIE-TC}.

\section*{Acknowledgments}

This work has been sponsored by the Irish Research Council, through individual PhD scholarships (LPJG and MC) and the Advance Laureate Award, IRCLA/2019/127. Additional support has been provided by the Science Foundation Ireland AMBER centre (12/RC/2278$_-$P2). The database collection at UC Davis was supported by the Critical Materials Institute, an Energy Innovation Hub funded by the U.S. Department of Energy, Office of Energy Efficiency and Renewable Energy, Advanced Manufacturing Office. We thank J. K. Byland, Shaoqing Ding, Rogelio Mata, Haley E. Magliari, Maxwell G. Wright for the data collection. The authors also thank Rossa Brennan and Emma Hughes for their help in annotating the training data for the $T_\mathrm{C}$ NER model.  We acknowledge the DJEI/DES/SFI/HEA Irish Centre for High-End Computing (ICHEC) and Trinity Centre for High-Performance Computing (TCHPC) for the provision of computational resources. 

\section*{Author Contributions}

L.P.J.G. and M.C. both contributed to the design and implementation of the pipeline and the gathering of training data for each of the models. M.C. trained the RF screening models for $T_{\mathrm{C}}$. V.T. provided part of the manually extracted $T_\mathrm{C}$ database and a set of relevant PDFs used for the extraction. S.S. conceived the idea for the project and supervised the overall research. L.P.J.G., M.C. and S.S. composed the manuscript.

\section*{Competing Interests}

The authors declare no competing interests.

\bibliographystyle{unsrt}
\bibliography{references-new}

\pagebreak

\listoffigures
\vfill
\end{document}